# Mapping AI Arguments in Journalism and Communication Studies

**Gregory Gondwe**

Department of Communication, California State University – San Bernardino, USA

**Abstract:** *This study investigates and suggests typologies for examining Artificial Intelligence (AI) within the domains of journalism and mass communication research. We aim to elucidate the seven distinct subfields of AI, which encompass machine learning, natural language processing (NLP), speech recognition, expert systems, planning, scheduling, optimization, robotics, and computer vision, through the provision of concrete examples and practical applications. The primary objective is to devise a structured framework that can help AI researchers in the field of journalism. By comprehending the operational principles of each subfield, scholars can enhance their ability to focus on a specific facet when analyzing a particular research topic.*

**Introduction**

Artificial Intelligence (AI) has emerged as a transformative force in journalism and mass communication research, revolutionizing how information is produced, disseminated, and consumed. As AI applications continue to proliferate in the media landscape, understanding its various subfields becomes crucial for researchers seeking to comprehend its impact and potential in the domain of journalism studies. For example, most scholars have explored the pivotal role of machine learning in the context of journalism. Dunham and Dunham (2020), investigated the application of AI-generated content in spinning narratives, revealing its potential to produce convincing misinformation at scale. On the other hand, natural language processing (NLP) has been employed to analyze sentiment in news articles, enabling researchers to gauge public reactions to specific topics (Broussard, et. al., 2019). Meanwhile, robotics has implications for automated news reporting, exemplified by the emergence of AI-powered chatbots delivering real-time news updates (Cheng, Wu, Li, & Jiang, 2020).

Given the incessant nature, this study delves into the realm of AI in journalism and mass communication research, with a specific focus on proposing typologies to facilitate comprehensive investigations. The primary objective is to present a framework that aids researchers in studying AI's multifaceted presence within the field of journalism. We begin by expounding on the seven fundamental subfields of AI: machine learning, natural language processing (NLP), speech recognition, expert systems, planning, scheduling, and optimization, robotics, and computer vision. Through specific examples and practical applications, we illustrate how each subfield operates within the context of journalism and mass communication. The analysis aims to offer a systematic approach for media and Communication scholars to navigate the diverse facets of AI within their research inquiries. By understanding the distinct characteristics and functionalities of each subfield, researchers can effectively focus their investigations on specific AI applications that align with their study's objectives. For instance, understanding the nuances of natural language processing (NLP) can enable researchers to analyze AI-generated content's impact on media consumption, while insights into robotics can shed light on the implications of automation in news reporting. Moreover, this study highlights the practical implications of AI typologies in journalism studies. By uncovering AI's role in journalism, scholars can gain deeper insights into the implications of automated news production, personalized content delivery, and the challenges posed by AI-generated misinformation. Additionally, the framework facilitates the identification of gaps and areas for further exploration, encouraging researchers to delve into specific AI subfields to address emerging challenges and opportunities.

Overarchingly, this study seeks to contribute to the expanding body of knowledge on AI in journalism and mass communication research by providing a structured and accessible framework. Through the proposed typologies and in-depth examination of AI's subfields, we aim to empower scholars to undertake informed and focused investigations into the multifaceted interactions between AI and the media landscape. Ultimately, a comprehensive understanding of AI's influence in journalism studies is critical for anticipating its future trajectory and guiding ethical and responsible AI deployment within the media industry.

**Arguments Surrounding AI in journalism Studies**

As AI continues to evolve, scholars in journalism grapple with the fluidity of arguments, seeking a solid framework to anchor their research. Nevertheless, when exploring AI in journalism studies, three main themes emerge as significant: Firstly, the concept of AI enhancing efficiency through automation stands out. This theme emphasizes the potential benefits and challenges of AI in tasks like content curation, personalized recommendations, sentiment analysis, and automated transcription and translation. Many scholars assert that automating journalistic practices allows journalists to save time and resources, enabling them to focus on more in-depth and investigative reporting (Canavilhas, 2022; Deuze & Beckett, 2022; Moran & Shaikh, 2022; Pavlik, 2023).

Secondly, the theme of quality and accuracy concerns in AI-generated content surfaces as a central debate. While AI can aid in fact-checking and content moderation, concerns arise about the reliability of AI-generated content. Language models like GPT-3 have demonstrated impressive capabilities in generating human-like text, but they may also produce misinformation or biased content. Mhlambi and Tiribelli (2023) argue that AI tools are inevitably designed with the concept of rationality, as opposed to relationality. This in turn, perpetuate the old tendencies. Gondwe's (2023) empirical study provides evidence supporting this argument, highlighting how certain biases may persist even when efforts are made to combat stereotypes using anecdotal data or metadata. Thus, rigorous editorial oversight is crucial to ensure the dissemination of accurate information when employing AI-generated content.

    Thirdly, AI's ethical considerations and societal implications come to the forefront. Mhlambi (2020) observes that AI's challenge lies in being centered on Western values, which are often driven by capitalism, conflicting with societal ethical needs. Criticisms such as the exclusion of marginalized communities in automated decision-making system design, biases in data selection, failure to recognize societal interconnectedness, the commodification of digital selves, and data centralization, raised by Couldry and Meijias (2020) underscore the capitalistic nature of the foundations that inform AI. As AI integrates into journalism, ethical concerns arise regarding the perpetuation of biases and stereotypes through AI tools trained on biased data. Moreover, debates focus on the impact of AI-driven content personalization on media diversity and its potential to create echo chambers where users receive content aligned with their existing beliefs. Ethical considerations also encompass issues like deepfakes and the manipulation of visual content that could mislead audiences. To comprehend these complexities and more, understanding the taxonomy of AI is essential.

**The Taxonomy of Generative AI**

    In a recent study, Gozalo-Brizuela and Garrido-Merchan developed a taxonomy of generative AI to represent the main mappings between a multimedia input and the output type of data. Accordingly, 9 categories emerged as in figure 1 below, namely, 1. Text-to-image models (whose input is a text prompt and an image output); 2. Text-to-3D models (involves the mapping of text prompts to dimensional images); 3. Text-to-Audio models (models whose input information is text and whose output information is audio); 4.Text-to-Video models (that generates images from texts); 5. Text-to-Text models (converts text into another text to satisfy tasks); 6. Text-to-Code models (converting an idea into code, i.e., with Codex); 7. Text-to-Science models (using AI to generating automatic scientific texts, i.e., as with the Galactica and Minerva models); 8. Human Emotion Diffusion model (models capable of generating human emotions); and 9. Models that generate and/or discover algorithms (i.e., Alpha Tensor that creates a more efficient algorithm for matrix multiplication relevant for improving the efficiency of algorithms. Figure 1

below presents a summary of the categories.

**Figure 1.** Nine Categories of Generative AI models

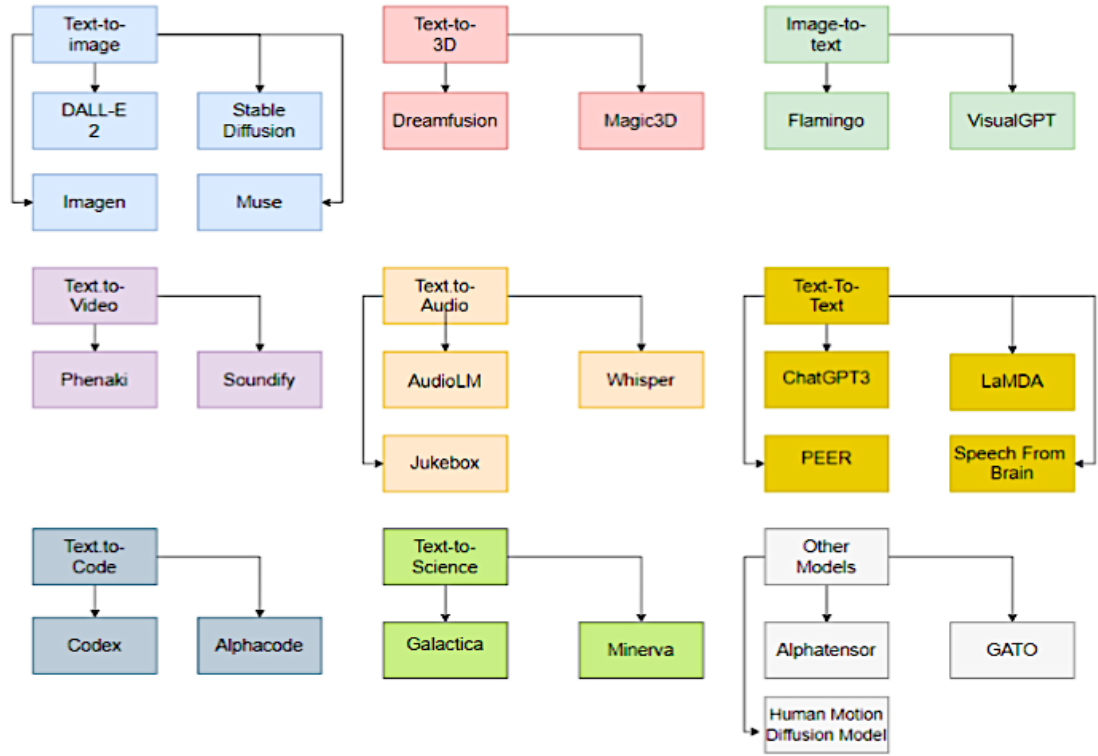

*"A taxonomy of the most popular generative AI model that have recently appeared classified according to their input and generated format". Source: Roberto Gozalo-Brizuela, Eduardo C. Garrido-Merchán, 2023, p. 3.*

**Generative AI in Journalism and Communication Studies: A Multidimensional Exploration**

The realm of generative AI, while encompassing a spectrum of possibilities, finds its unique resonance within the domains of journalism and communication studies. However, it's worth noting that while various dimensions of generative AI hold promise in this field, only a select few have been integrated into the practical aspects of news writing and reporting. This asymmetry, as illuminated by Gozalo-Brizuela and Garrido-Merchán (2023), may partly stem from the ongoing development of many AI subfields and their gradual adoption in diverse academic disciplines and industries. These subfields are better summarized in the figure below.

**Figure 1:** The seven fields of AI

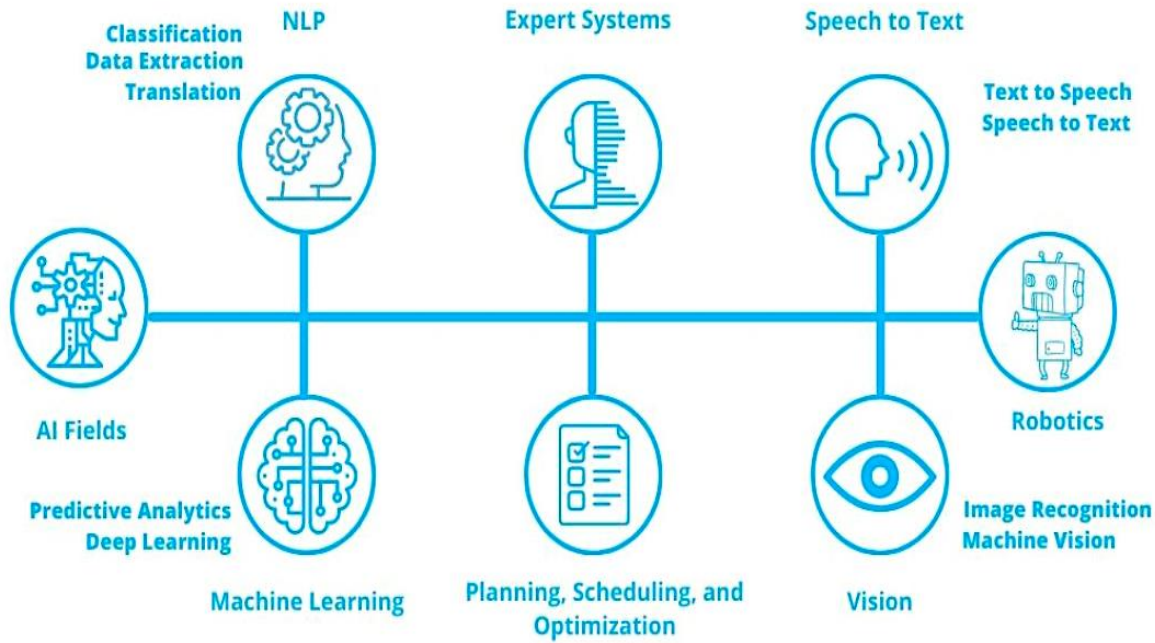

**Source:** *de-Lima-Santos, Mathias-Felipe, and Wilson Ceron. 2022. Artificial Intelligence in News Media: Current Perceptions and Future Outlook. Journalism and Media3: 13–26. https://doi.org/10.3390/*

To appreciate the intricate relationship between generative AI and journalism and communication studies, it is illuminating to examine the seven distinctive dimensions or subfields within which generative AI operates:

**1. Machine Learning**
Machine learning involves designing algorithms that build models from data without pre-existing solutions to the problem, enables the tool to improve performance based on previous experience (Russell and Norvig, 2021, p. 19). Machine learning can be utilized by journalists to transcribe or translate information. Machine learning, as a subfield of Artificial Intelligence (AI), offers significant opportunities for application within Journalism and Communication studies. Its capabilities in data analysis, pattern recognition, and predictive modeling can revolutionize various aspects of media, journalism, and communication research.

Among the many other things to which journalists or media practitioners can employ machine learning is in Automated News Writing (ANW). Through this process, the software is trained to generate news articles based on structured data, such as financial reports or sports statistics. Generating automated news writing using machine learning involves training models to understand the structure of news articles, identify relevant information, and generate coherent and contextually accurate content. Below is a step-by-step overview of the process with practical examples:

**Step1: Data Collection:** A journalist can gather a large dataset of news articles from various reliable sources, i.e., news stories of the 2023 Women Soccer -FIFA World Cup. The dataset should include labeled examples of different article types, such as each group's standing, predictions, and sometimes other stories emerging from the competition, i.e., health, other forms of entertainment outside the played, etc. FIFA, BBC, Reuters, NYT, and many others can serve as reliable news sources.

**Step2: Data Preprocessing:** The second step would involve the cleaning and pre-processing of your data. Cleaning and preprocessing the data involves the removing of noise by formatting inconsistencies, and irrelevant information. This step may involve tokenization, stemming, and removing stop words. Several methods can be applied depending on the kind of data. In fact, generative AI can also aid this process. However, we caution that journalists use the chatbots for such works sparingly. Reliability pretests should be conducted. Once the dataset is processed, journalists should split it into training and testing sets.

**Step3: Feature Engineering:** In the third step, journalists will Convert the text data into numerical features that the machine learning model can process. This can be achieved using techniques like TF-IDF (Term Frequency-Inverse Document Frequency) or word embeddings like Word2Vec or GloVe. Here is one simple way of using TF-IDF to turn words in news articles into numbers for the computer can understand them better.

Imagine we have 10 news articles on Women Soccer. First, we count how many times each word appears in each article. Then, we figure out how important each word is in all the articles together. For example, the word "the" appears in all 10 articles, so it's not very important because it's common. But the word "hat-trick" appears only once, so it's more important because it's rare. TF-IDF gives each word a special number that shows how important it is in each article. This helps computers understand the main topics in the news articles and find similarities between them to help autogenerate the news story. In the end, the news articles are represented as numerical values, making it easier for computers to analyze and process the information they contain. This whole process can be performed under Machine Learning, particularly NLP.

**Step 4: Model Selection and Training, Generating news content and Evaluation:**

Having categorized the dataset and identified the central thrust, journalists chose the most appropriate machine learning model for automated news writing. Commonly used for natural language generation tasks are Recurrent Neural Networks (RNNs) and Long Short-Term Memory (LSTM) networks. Alternatively, transformer-based models like GPT-3 can also be explored for more advanced language generation capabilities.

The transformer model, such as GPT-3, has been previously pre-trained on a massive amount of data to learn language patterns. The role of a journalist, at this stage would involve providing the model with additional data related to the news and guiding it to optimize the generation of coherent and informative news stories. Therefore, a journalist would need to provide a specific input prompt to the model. The prompt can be a headline, a summary, or a few sentences about the topic. For instance, a prompt like "Japan is predicted to win the 2023 World Cup Championship..." is given, prompting the model to complete the sentence and generate the rest of the article. It is recommended to provide as many details as possible, given the presence of a categorized dataset.

After inputting the prompt, the model generates the rest of the news story based on its understanding of the language and patterns it has learned during training. The generated text might require some post-processing to improve coherence, and structure, and remove any grammatical errors or inaccuracies. The generated content is post-processed to ensure coherence and readability. The generated articles are also evaluated against human-written articles to assess the quality and accuracy of the automated news writing. It is acknowledged that automated news stories generated by the model might not always be perfect, necessitating human editors to review the output, make necessary corrections, and ensure the story is factually accurate and meets journalistic standards before publishing.

**Practical Example:** Let's assume we have a trained model for generating sports news articles. We provide the prompt: "In an exciting 2023 Women's World Cup match last night, Japan secured a thrilling 5 – 0 victory over their Zambia..." to the model. The model generates the following complete news article:

*Japan Secures Thrilling 5-0 Victory Over Zambia in 2023 Women's World Cup Clash*

*In a riveting and action-packed match at the 2023 Women's World Cup, Japan showcased their prowess on the soccer field, securing a resounding 5-0 victory over their Zambia counterparts. The thrilling encounter took place last night at the stadium, and fans were treated to an impressive display of skill and determination from both teams.*

*From the opening whistle, it was clear that Japan meant business, as they quickly took control of the game. Their offensive tactics and seamless teamwork allowed them to make continuous advances towards the Zambian goal, putting immense pressure on their*

*defense. The Japanese players' precise passing and swift footwork left the Zambian defenders struggling to keep up with their relentless attacks.*

*As the first half progressed, it became evident that Japan's strategy was paying off. In the 25th minute, Japan's star striker, Aiko Tanaka, skillfully maneuvered past the Zambian defenders and unleashed a powerful shot, finding the back of the net with remarkable accuracy. The crowd erupted in cheers as Japan took the lead.*

*However, the Japanese team was not content with a mere one-goal advantage. They continued to press forward with unwavering determination, exploiting gaps in the Zambian defense. In the 38th minute, Yuki Suzuki capitalized on a defensive error and scored Japan's second goal, extending their lead further.*

*The second half witnessed an even more dominant performance from Japan. Their players demonstrated exceptional ball control and tactical awareness, leaving the Zambian defense struggling to contain their attacking force. In the 58th minute, Tomoko Nakamura delivered a stunning header from a perfectly executed corner kick, further widening the gap to 3-0.*

*Japan's relentless assault on the Zambian goal continued, with each player contributing to the team's remarkable performance. In the 71st minute, midfielder Risa Yamamoto fired a powerful shot from outside the penalty box, sending the ball past the Zambian goalkeeper and into the net. The fourth goal sparked jubilant celebrations among the Japanese players and their supporters.*

*The match's final moments were marked by a remarkable display of sportsmanship and teamwork from both sides. However, it was Japan who had the last word, with Aiko Tanaka scoring her second goal of the night in the 87th minute, sealing a phenomenal 5-0 victory.*

*Japan's coach, Yuki Nakamura, expressed immense pride in his team's performance. "The players executed the game plan brilliantly, and their dedication and hard work were evident throughout the match," he said in a post-match interview.*

*As the final whistle blew, the stadium erupted in cheers and applause for Japan's remarkable victory. The team's outstanding display on the field has undoubtedly made them a formidable force in the 2023 Women's World Cup. With such a commanding performance, Japan has sent a clear message to other opponents that they are a team to watch out for in the tournament.*

*"In an exciting match last night, the home team secured a thrilling victory over their rivals. The match was marked by intense competition, with both teams displaying exceptional skills and teamwork. The home team's star striker scored a hat-trick, leading the team to a 3-2 victory in the dying minutes of the game. The crowd erupted with joy as they witnessed their favorite players' outstanding performance. The coach praised the team's determination and spirit, emphasizing the hard work put in during the training sessions. This victory cements the Japan's position at the top of the league table, further fueling their aspirations for the championship title."*

Although the above news story appears comprehensive, its veracity is subject to scrutiny (Gondwe, 2023). This is due to the model's dependence on the extent of training it has received (Pavlik, 2023). Consequently, the story remains unsuitable for dissemination until it undergoes rigorous fact-checking and human editorial review (Deuze & Beckett, 2022).

Additionally, journalists can leverage machine learning techniques for a wide array of applications, encompassing personalized content recommendations, sentiment analysis, fact-checking and fake news detection, audience analysis, automated transcription and translation, image and video analysis, predictive analytics, automated content moderation, and media recommendation systems (Moran & Shaikh, 2022). These diverse applications offer valuable tools and insights to enhance journalistic practices and cater to the evolving demands of the media landscape (Appelgren and Nygren, 2014).

With personalized content recommendations, machine learning algorithms can analyze user behavior, preferences, and historical interactions to deliver tailored news and information to individual users (de-lima-Santos & Ceron, 2021). Sentiment analysis enables journalists to gauge public sentiments and reactions towards specific topics, allowing them to better understand the audience's response to news stories and adapt their reporting accordingly (Pavlik, 2023)). Furthermore, fact-checking and fake news detection algorithms play a crucial role in verifying the authenticity of information, enhancing the credibility of journalistic content (Broussard, et. al., 2019). Audience analysis powered by machine learning enables journalists to gain deeper insights into their readership, understanding their interests, demographics, and engagement patterns. Automated transcription and translation facilitate efficient and accurate language processing, enabling journalists to cover news from diverse linguistic regions seamlessly. Additionally, machine learning-based image and video analysis tools aid in identifying and categorizing visual content, enhancing multimedia storytelling and investigative reporting (Appelgren and Nygren, 2014).

Predictive analytics can be harnessed to forecast news trends and identify potential newsworthy events, allowing journalists to stay ahead in the rapidly changing news landscape (Appelgren and Nygren, 2014). Automated content moderation mechanisms assist in maintaining the quality and appropriateness of user-generated content in digital media platforms, fostering a more constructive online community (Gondwe, Ferrucci, & Tandoc, 2022). Lastly, media recommendation systems leverage machine learning algorithms to suggest relevant content to users based on their preferences, enhancing user engagement and fostering a more personalized media consumption experience (Wanda, et. al., 2021). The integration of these machine learning applications empowers journalists with data-driven insights and efficient tools, ultimately enriching the journalistic process and serving the informational needs of the global audience (Harvard, 2020).

2. **Computer Vision (CV)**

Computer vision presents a powerful set of tools for journalists to enhance their storytelling, streamline data analysis, and improve the accuracy and impact of their reporting. This field allows computers to derive meaningful information from digital images (Marconi, 2020) such as facial recognition and other forms of image interpretation. Therefore, journalists can utilize computer vision technology to enhance their reporting and storytelling in various ways, including:

Journalists can leverage computer vision to enhance their image and video analysis capabilities, facilitating more efficient interpretation of visual content (Marconi, 2022). Image recognition algorithms enable the identification of objects, individuals, or locations in photos, enabling swift content categorization and organization (Gondwe and White, 2022. Likewise, video analysis can pinpoint critical moments, uncover patterns, and extract relevant information from the footage, bolstering the depth and accuracy of journalistic reporting (Wanda, et. al., 2021).

One specific example is the use of computer vision for fact-checking and verification of visual content, such as images and videos, in news articles (de-Lima-Santos & Ceron, 2021; Gondwe, 2018).). Here's how it works:

1. *Verification of User-Generated Content (UGC): With* the rise of social media and citizen journalism, journalists often come across user-generated content shared on platforms like Twitter, Instagram, and YouTube (Gondwe, 2023). This content may include photos and videos claiming to depict certain events or incidents.

2. *Reverse Image Search:* Journalists can use computer vision tools to perform reverse image searches. By uploading an image or providing a URL, these tools can quickly scan the internet to find similar or identical images. This helps in verifying the authenticity of the image and identifying its original source.

3. *Metadata Analysis:* Computer vision can also extract and analyze metadata embedded in images. Metadata includes information like the date, time, and location where the image was taken. Journalists can cross-reference this information with the claims made in the content to determine its accuracy.

4. *Image Forensics:* Computer vision algorithms can detect signs of image manipulation or tampering. This includes identifying edited areas, alterations in lighting, or the addition of objects that weren't present in the original scene.

5. *Deepfake Detection:* Deepfakes are highly realistic AI-generated videos that can deceive viewers. Computer vision models are being developed to detect signs of deepfake manipulation in videos by analyzing facial movements, audio discrepancies, and other visual cues.

6. *Contextual Analysis:* Computer vision can help journalists analyze the context in which an image or video was captured. This involves identifying landmarks, objects, or specific individuals in the visual content to provide additional context to the story.

7. *Real-Time Monitoring:* Some computer vision systems can monitor live streams and social media feeds for visual content related to breaking news events. This enables journalists to quickly identify and verify emerging stories.

8. *Data Visualization:* Journalists can use computer vision to create data visualizations and infographics from raw data. This can make complex information more accessible to readers and enhance the storytelling aspect of journalism.

Another practical example is the incorporation of augmented reality (AR) technology, allowing journalists to introduce interactive elements to their narratives. For instance, a reporter can offer an immersive and captivating experience to the audience by superimposing digital content onto real-world scenes. Stories dealing with historical information, visualizations, or animated graphics can be seamlessly integrated into physical locations, enriching the storytelling process. One such example is that of the New York Times visualization of Usain Bolt in comparison to 116 years of sprinters.

**Figure 2:** A complicated story made simple and awesome.

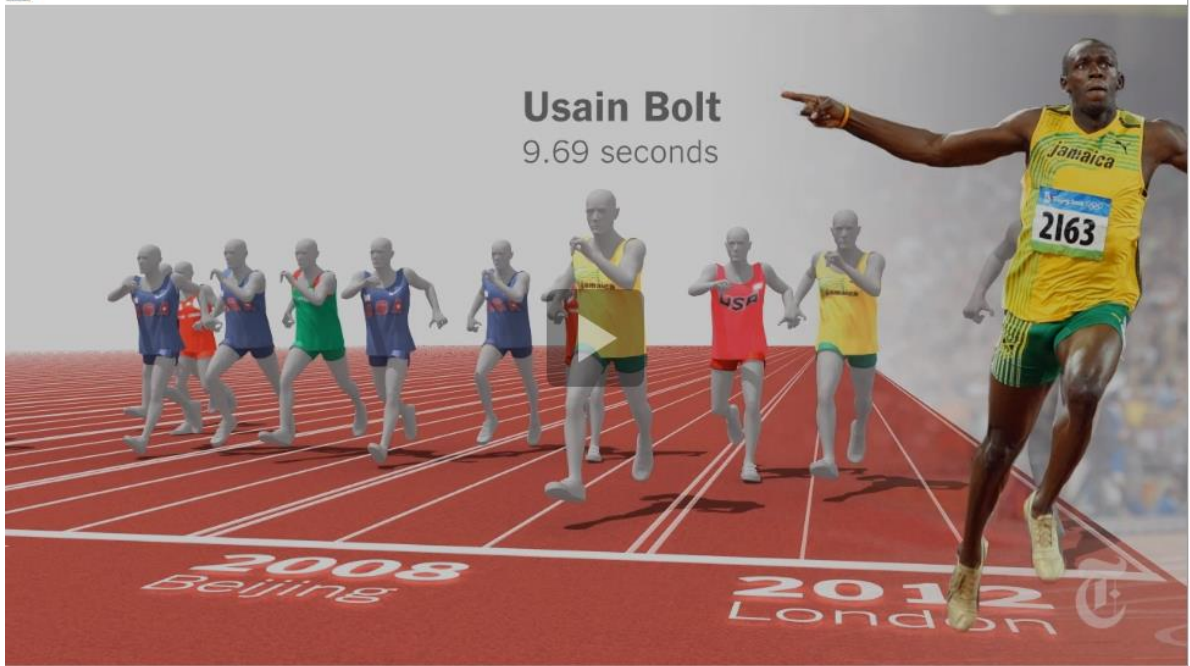

Source: https://archive.nytimes.com/www.nytimes.com/interactive/2012/08/05/sports/olympics/the-100-meter-dash-one-race-every-medalist-ever.html, and http://www.cjr.org/data_points/spreadsheets_olympics.php?page=all

Computer vision technologies offer journalists a multitude of capabilities, ranging from automating content moderation on online platforms to enhancing visual data analysis. By scrutinizing images and videos, journalists can detect inappropriate or harmful content, actively contributing to a safer and more respectful online environment. This empowers media organizations to maintain the quality and appropriateness of user-generated content in digital media platforms.

Computer vision aids in visual data analysis, enabling journalists to analyze extensive visual datasets, such as satellite images or charts, uncovering nuanced trends and insights that might elude human observation. This ability to process large volumes of visual data enhances the depth and accuracy of journalistic reporting). Consequently, journalists can present complex information more effectively through data visualization, generating interactive charts, graphs, and infographics to enhance accessibility and audience engagement.

Facial recognition algorithms facilitate the identification of individuals in visual content, augmenting the journalistic capacity for human subjects' identification (Marconi, 2020). Emotion analysis further deepens understanding by gauging emotional responses, offering profound insights into how individuals react to specific events or circumstances (Dunham & Dunham, 2020). Additionally, computer vision plays a pivotal role in fact-checking processes. By analyzing metadata, detecting manipulations, and cross-referencing with reputable sources, journalists ensure the authenticity and accuracy of visual content within their reports (Harvard, 2020). Embracing these computer vision applications, journalists can elevate their storytelling and investigative capacities, enriching news reporting with enhanced visual understanding and meaningful context (de-Lima-Santos & Ceron, 2021).

3. **Speech Recognition,**
   Speech recognition technology, focused on automatically transcribing human speech into text, finds applications in voice-activated commands and question-answering systems (de-Lima-Santos & Ceron, 2021). Journalists can harness this technology to streamline their reporting processes and deliver content more effectively. Here are expanded examples of how speech recognition benefits journalism with additional citations:

**Transcription Services**:
- *Example*: Journalists can employ speech recognition software to transcribe interviews, saving substantial time and enabling them to focus on content analysis.
- *Benefit*: This not only reduces the time spent on transcription but also allows for deeper analysis and summarization of the conversation.

**Voice-to-Text Notetaking**:
- *Example*: While covering events or press conferences, journalists can use speech recognition apps on smartphones for real-time conversion of spoken notes into text.
- *Benefit*: This ensures accurate capture of essential details and facilitates easy reference during the writing process.

**Interview Transcripts for Articles**:
- *Example*: Journalists can transcribe interviews using speech recognition and seamlessly integrate relevant quotes into their articles.
- *Benefit*: It streamlines the process of quote extraction, enhancing the credibility and engagement of articles.

**Voice-Activated Assistants for Research**:
- *Example*: Voice-activated virtual assistants like Siri and Google Assistant offer journalists quick access to information, aiding in research.
- *Benefit*: Instant data retrieval keeps journalists informed on the go.

**Dictation for First Drafts**:
- *Example*: Speech recognition software can be used to dictate initial article drafts, accelerating the writing process.
- *Benefit*: It helps overcome writer's block and promotes free-flowing thoughts, with later refinement and editing.

**Multilingual Reporting**:
- *Example*: Speech recognition aids in the translation and transcription of interviews in different languages for international reporting.
- *Benefit*: It allows journalists to bridge language barriers and

maintain accurate global event reporting.

**Podcast and Video Transcript Generation**:
- *Example*: Speech recognition generates transcripts for multimedia content, enhancing accessibility.
- *Benefit*: Transcripts make multimedia content more accessible and searchable.

**Real-Time Reporting**:
- *Example*: Real-time transcription of live events using speech recognition helps journalists quote speakers accurately.
- *Benefit*: Timely reporting reduces the risk of misquoting or misinterpretation.

The integration of speech recognition technology into journalism workflows has a transformative impact on the profession. It empowers journalists with newfound efficiency, heightens the accuracy of content, and ultimately elevates the overall quality of reporting. These advantages align with the evolving demands of the media landscape and the expectations of today's information-hungry audiences, making speech recognition an invaluable tool for modern journalism.

**4. Natural Language Processing (NLP)**

Natural Language Processing (NLP) stands as an intricate subfield of computational linguistics that extends the capabilities of computer programs to engage with human language. This transformative discipline, as elucidated by Castro and New (2016), empowers machines to comprehend and respond to text and voice data, extract meaning from sentences, and generate coherent textual content. NLP encompasses a rich tapestry of applications, including translation, classification, clustering, and information extraction, while Natural Language Generation (NLG) and Natural Language Understanding (NLU) orbit closely in its lexicon.

For journalists, NLP serves as an invaluable toolbox for elevating news reporting to new heights. Its multifaceted applications are transformative in their own right, offering journalists a myriad of opportunities to refine and augment their craft:

**Sentiment Analysis**:
- **Example**: Journalists can employ sentiment analysis tools to decipher public sentiment concerning specific subjects or events, such as gauging sentiment towards a political candidate through social media posts.
- **Benefit**: This aids journalists in comprehending public opinion, informing their reporting, and unveiling trends and controversies.

**Automated Content Summarization**:
- **Example**: NLP algorithms can automate the generation of concise summaries for voluminous reports, speeches, or documents, allowing journalists to distill complex information efficiently.
- **Benefit**: This time-saving tool empowers journalists to swiftly access vital information, enhancing the efficiency of their reporting.

**Named Entity Recognition (NER)**:

- **Example**: NER helps identify and categorize entities like individuals, organizations, and locations mentioned in news articles, facilitating the identification of key players in a story.
- **Benefit**: Journalists can cross-reference data, fact-check claims, and ensure reporting accuracy through NER.

**Language Translation**:
- **Example**: NLP-driven translation tools assist journalists in converting interviews, articles, or statements from one language to another, broadening the spectrum of accessible sources.
- **Benefit**: This enables comprehensive global event coverage and broadens the audience reach.

**Topic Modeling**:
- **Example**: Journalists employ topic modeling to unveil prevalent themes within extensive text corpora, illuminating trending topics like Twitter discussions during elections.
- **Benefit**: It aids in spotting emerging trends, tracking public discourse, and steering reporting towards pertinent and timely matters.

**Fact-Checking and Verification**:
- **Example**: NLP tools assist journalists in fact-checking assertions made by public figures or in user-generated content, upholding the credibility of news reporting.
- **Benefit**: Fact-checking with NLP ensures readers receive accurate information.

**Automated News Generation**:
- **Example**: NLP algorithms generate automated news reports, particularly for data-driven stories, like sports scores and financial earnings reports.
- **Benefit**: Automation streamlines routine reporting, allowing journalists to dedicate more time to in-depth and investigative journalism.

**Social Media Monitoring**:
- **Example**: NLP tools monitor social media conversations surrounding specific events or issues, facilitating the collection of real-time information and eyewitness accounts.
- **Benefit**: It enhances the journalist's ability to amass real-time data and eyewitness testimonies, contributing to comprehensive reporting.

**Content Recommendation**:
- **Example**: News websites deploy NLP-powered recommendation systems to suggest related articles based on reader interests and history.
- **Benefit**: These systems keep readers engaged, promoting exploration of diverse news topics.

By weaving NLP into their reporting tapestry, journalists forge a path toward heightened accuracy, efficiency, and journalistic depth while staying attuned to public sentiment and emergent trends. This symbiotic relationship between journalism and NLP promises an ever-evolving landscape of news reporting and consumption.

5. **Planning, Scheduling, and Optimization**,
Planning, Scheduling, and Optimization which are systems that use AI to determine steps to take (planning) and understand when to carry out a certain step (scheduling) to achieve a goal. These systems can also find the most optimal way to make all necessary adjustments in the most efficient manner (optimization). In the dynamic world of journalism and news reporting, the strategic application of planning, scheduling, and optimization principles is pivotal to the seamless and impactful flow of news content (Coddington, 2015, Stray, 2021). These principles serve as the guiding compass for news organizations, ensuring they navigate the complex landscape of information dissemination effectively. Let's explore these concepts in greater detail with real-world examples:

**Editorial Planning:** Editorial planning is the foundation of news reporting. Newsrooms meticulously plan their coverage by identifying important events, stories, and topics to address. This involves considering the news value, target audience, available resources, and editorial priorities (Reese & Shoemaker, 2018).
*Example:* Suppose a newsroom plans to cover a significant political event, such as a presidential election. They will allocate resources to gather information from polling stations, interview candidates and voters, and prepare in-depth analysis. This planning ensures comprehensive and balanced election coverage.

**Content Scheduling:** Content scheduling is about timing the release of news stories to maximize their impact. News organizations consider when their audience is most active and receptive to news updates (Reese & Shoemaker, 2018).
*Example:* An online news portal schedules breaking news stories to coincide with peak traffic hours. This may involve releasing stories during morning and evening commutes or around lunchtime to reach a larger audience.

**Resource Allocation:** Effective resource allocation involves deciding how to distribute reporters, photographers, videographers, and other resources to cover various stories. It ensures that newsrooms can cover multiple events simultaneously and meet deadlines
*Example:* A news agency planning to cover both a natural disaster and a political rally on the same day will allocate resources strategically. They may assign experienced reporters to the disaster and assign political experts to the rally for in-depth coverage.

**Optimized Content Distribution:** Content distribution optimization leverages data analytics and algorithms to tailor content delivery to the preferences and behaviors of the audience. Factors like user demographics, geographic location, and content engagement are considered.
*Example:* An online news aggregator app optimizes content recommendations by analyzing user behavior. If a user frequently reads science articles, the app's algorithm will prioritize recommending more science-related content to enhance user engagement.

**Production Workflow:** Within news production workflows, scheduling ensures that tasks are efficiently coordinated to meet deadlines. It involves defining timelines for research, writing, editing, fact-checking, and design.
*Example:* In a television newsroom, the production schedule for a live broadcast is meticulously planned. This schedule allocates time for reporters to gather information, write scripts, rehearse, and ensure that everything runs smoothly during the live show.
**Resource Efficiency:** Resource efficiency optimization aims to reduce costs while maintaining quality. News organizations often automate repetitive tasks and share content across multiple platforms to reach a wider audience.
*Example:* Automated transcription software is used to transcribe interviews and speeches quickly and accurately, reducing the time and cost associated with manual transcription.

**Social Media Posting:** social media plays a pivotal role in news distribution. News outlets schedule posts to coincide with peak social media activity, ensuring that breaking news and updates reach a broader audience.
*Example:* During a live event like a press conference, a news organization schedules tweets and Facebook posts in real-time to keep the audience informed and engaged as the event unfolds.

**Ad Placement Optimization:** Online journalism relies on ad placement optimization to maximize revenue. Algorithms analyze user behavior and content relevance to strategically place ads within web pages or mobile apps.
*Example:* An online news website optimizes ad placement to increase ad click-through rates. For instance, it may display fashion-related ads to users who frequently read fashion news, resulting in higher ad revenue.

These detailed explanations illustrate how planning, scheduling, and optimization are indispensable tools in the journalist's toolbox. They help streamline news production, enhance audience engagement, and ensure that news organizations can deliver timely, relevant, and efficient news reporting. These principles are not only essential for journalism but also fundamental for maintaining the credibility and trustworthiness of news in the digital age.

6. **Expert Systems,**
Expert Systems are computer systems that emulate the problem-solving and decision-making capabilities of human experts in specific domains. They are designed to capture and utilize the knowledge and expertise of human specialists to provide intelligent solutions and recommendations. Expert Systems are built using knowledge representation techniques, inference engines, and a knowledge base. The knowledge base contains domain-specific information, rules, and heuristics that represent the expertise of human specialists. The inference engine applies logical reasoning and inference mechanisms to process the knowledge base and provide responses or solutions to user queries or problems. It is important to note that Expert Systems heavily rely on explicit knowledge provided by human experts and may struggle with uncertainty, ambiguity, and incomplete information. They are also less adept at learning and adapting to new situations without human intervention. In short, Expert systems use databases of knowledge and rules to solve complex problems (de-Lima-Santos & Ceron, 2021; Russell and Norvig, 2021).

Expert systems, a branch of artificial intelligence (AI), have found applications in journalism and news reporting, enhancing various aspects of the journalistic process. Here's how expert systems are used in this field:

**Automated Fact-Checking**:
- *Explanation*: Expert systems designed for fact-checking analyze news articles and compare the claims made within them with a database of credible sources and historical data. When inconsistencies or inaccuracies are detected, these systems flag the content for further review by human editors.
- *Example*: Full Fact, a UK-based fact-checking organization, employs automated fact-checking systems that use NLP and machine learning to identify and verify claims in news articles and political speeches.

**Content Recommendation**:
- *Explanation*: Expert systems analyze user data, including reading habits, preferences, and behavior, to suggest relevant news articles. Algorithms learn from user interactions and tailor content recommendations to maximize user engagement.
- *Example*: Netflix's content recommendation system is a prime example. While not focused on news, its algorithm analyzes user viewing habits to suggest movies and TV shows, creating a personalized viewing experience.

**Automated News Generation**:
- *Explanation*: Automated news generation systems use structured data inputs to generate news articles with minimal human intervention. These systems are particularly useful for data-driven stories, where large datasets can be converted into readable news reports.
- *Example*: Automated Insights' Wordsmith platform generates financial news articles for the Associated Press. The platform transforms earnings reports from companies into written news stories, providing rapid coverage of financial events.

**Audience Analytics**:
- *Explanation*: Expert systems analyze audience data to gain insights into reader demographics, geographic

locations, and interests. This information informs editorial decisions, content strategies, and advertising targeting.
- *Example*: The New York Times uses audience analytics to understand its readers better, tailoring content and advertising to specific demographics and interests.

**Language Translation**:
- *Explanation*: Expert systems specializing in natural language processing (NLP) translate news articles from one language to another. This enables news organizations to provide content to international audiences.
- *Example*: Google Translate is a widely used NLP system that provides translations for news articles, making them accessible to readers around the world (Google, n.d.).

**Predictive Analytics**:
- *Explanation*: Expert systems leverage predictive analytics by analyzing historical data and trends to forecast future news events. Journalists can use these predictions to identify potential stories and stay ahead of emerging trends.
- *Example*: Bloomberg uses predictive analytics to forecast financial market trends and provide news articles on potential market movements, giving investors valuable insights.

**Social Media Monitoring**:
- *Explanation*: Expert systems continually monitor social media platforms for breaking news and trends. This real-time monitoring helps journalists identify and report on stories as they unfold.
- *Example*: Newsrooms use social media monitoring tools like CrowdTangle to track trending topics and breaking news on platforms like Facebook and Instagram (CrowdTangle, n.d.).

**Automated Research**:
- *Explanation*: Expert systems assist journalists in conducting research by rapidly sifting through vast amounts of data to identify relevant sources and information. This accelerates the information-gathering process.
- *Example*: The Washington Post's Heliograf uses AI to assist reporters in researching and writing stories, particularly for data-driven topics like election results and sports scores (The Washington Post, 2016).

Expert systems have become integral to modern journalism, helping news organizations produce timely, accurate, and relevant content while also improving the reader experience through personalization and data-driven insights. These systems enable journalists to focus on higher-level tasks like analysis and storytelling, while routine processes are handled efficiently by AI-powered tools.

## 7. Robotics,

Robotics, which integrates different cognitive technologies to enable computers and systems to perform different tasks in conjunction with people in unpredictable environments. Robotics has found applications in journalism practice, enhancing various aspects of news reporting and storytelling. Here are ways in which robotics are used in journalism:

**Automated Reporting**:
- *Explanation*: Robots equipped with natural language processing (NLP) capabilities can generate news reports based on structured data inputs. These robots are particularly useful for covering data-driven stories, such as financial earnings reports and sports scores.
- *Example*: The Associated Press (AP) uses a robotic journalist called Wordsmith, which transforms structured data into written news stories. It has been employed for generating earnings reports (AP, 2015).

**Data Gathering in Hazardous Environments**:
- *Explanation*: Robots can be deployed in dangerous or challenging environments where human journalists may face risks. They can gather visual and sensory data, providing real-time information from areas like conflict zones or natural disaster sites.
- *Example*: Drones equipped with cameras have been used to capture images and videos in conflict zones, enabling remote journalism without endangering reporters' lives (BBC, 2018).

**Camera Drones for Aerial Footage**:
- *Explanation*: Drones equipped with high-resolution cameras are employed for capturing aerial footage, offering unique perspectives in journalism. This is commonly used in documentary filmmaking and news coverage.
- *Example*: News organizations use camera drones to capture footage of events like wildfires, protests, and natural disasters, providing a comprehensive view of the situation from the air (CNN, 2014).

**Automated Photography**:
- *Explanation*: Robotic cameras and camera rigs are used to capture still images. These systems can be programmed to take photos at specific intervals or in response to certain triggers.
- *Example*: Some news outlets use automated camera systems for capturing time-

lapse photos of events like construction projects or changing weather conditions (New Atlas, 2017).

**Telepresence and Remote Reporting**:
- *Explanation*: Telepresence robots allow journalists to remotely participate in interviews or events. These robots can move, interact, and capture audiovisual content in real time.
- *Example*: Journalists have used telepresence robots to conduct interviews or attend press conferences remotely, reducing the need for physical presence (TechCrunch, 2013).

**Interactive Storytelling**:
- *Explanation*: Robots can be programmed to engage in interactive storytelling, such as answering questions or providing information through chatbots on news websites or social media platforms.
- *Example*: Chatbots like The Washington Post's "Heliograf" have been used to provide users with real-time updates on topics like sports scores and election results (The Washington Post, 2016).

**Robotic Video Production**:
- *Explanation*: Robots can assist in video production by handling tasks like camera operation, lighting adjustments, and even basic video editing.
- *Example*: Some news organizations use robotic camera systems to automate the capture of video footage, particularly for studio-based reporting (BBC News, 2010).

**Automated Video Captioning**:
- *Explanation*: Robots equipped with speech recognition capabilities can automatically generate captions for videos, enhancing accessibility for viewers.
- *Example*: Automated captioning systems are used by news outlets to ensure that videos are accessible to individuals with hearing impairments (BBC, 2020).

Robots and robotic technologies are increasingly integrated into journalism practice, enabling more efficient reporting, enhancing storytelling capabilities, and providing access to otherwise inaccessible or hazardous environments. While they are tools that assist journalists, human expertise remains essential in selecting, interpreting, and contextualizing the information gathered by these robotic systems.

*Conclusion: A bustling newsroom, filled with journalists, editors, and researchers*

As we venture into the digital age, the winds of change are sweeping through the world of journalism, and at the heart of this transformation stands a powerful ally - Artificial Intelligence (AI). In this chapter, we embark on an exciting journey to explore the captivating world of AI in journalism, dissecting the myriad benefits and intriguing implications that accompany its integration.

*Journalist A: "Efficiency and Speed"*
Imagine a newsroom where AI handles the mundane, time-consuming tasks - fact-checking, data analysis, and content generation. This newfound efficiency allows us, the storytellers, to channel our creative energies into deeper investigative work, enriching our narratives like never before.

*Journalist B: "Data-Driven Journalism"*
AI's knack for crunching numbers and spotting patterns in vast datasets opens a treasure chest of opportunities. We can now harness AI's analytical prowess to uncover stories hidden within the numbers, delivering more comprehensive, data-driven journalism to our readers.

*Journalist C: "Personalization"*
Our audience is diverse, each with unique tastes and preferences. With AI, we can tailor content to suit individual readers, crafting a personalized news experience that keeps them engaged and informed.

*Journalist D: "Enhanced Storytelling"*
AI isn't just about words on a page; it's about immersion. Virtual Reality (VR) and Augmented Reality (AR) journalism, powered by AI, catapult our storytelling to new heights. Readers don VR headsets to walk through the scene of a breaking news event, forging an emotional connection like never before.

*Journalist E: "Fact-Checking"*
In an era plagued by misinformation and fake news, AI emerges as a trusty sidekick. It can rapidly cross-reference information against reliable sources, battling falsehoods with the power of verification.

*Journalist F: "Ethical Conundrums"*
Ah, but let's not skip over the challenging part. With great power comes great responsibility. The ethical quandaries surrounding AI are as complex as the algorithms themselves. Bias, privacy concerns, and the potential for AI to deceive - these are weighty matters we must grapple with.

*Journalist G: "Job Displacement"*
And then there's the elephant in the room - job displacement. Are we at risk of trading our newsroom desks for algorithms? The automation of certain tasks could indeed reshape our profession.

*Journalist H: "Quality Control"*
Let's not forget quality control. If we lean too heavily on AI-generated content, will it erode the quality and trustworthiness of journalism? After all, there's no algorithm for integrity.

*Journalist I: "Algorithmic Bias"*
Last but not least, we confront the specter of algorithmic bias. AI is only as good as the data it's trained on. What if the data carries inherent biases? Could we inadvertently perpetuate prejudice in our reporting?

In this whirlwind discussion, we've journeyed through the promises and perils of AI in journalism. It's a transformative force, one that beckons us to innovate responsibly, embracing AI as a collaborator rather than a replacement. The future of journalism is intricately intertwined with AI, and it's up to us, the stewards of truth, to ensure it remains a force for good in the ever-evolving landscape of news reporting.